# Advances in Thermal Modeling of Selective Laser Sintering of Single-Component Metal Powders

## Bin Xiao[1], Yuwen Zhang[2]

1: New Technology Team, Tyco Fire Suppression & Building Products, USA, bxiao@tycoint.com
2: Department of Mechanical and Aerospace Engineering, University of Missouri, USA, zhangyu@missouri.edu

**Abstract** Selective laser sintering (SLS) of single component metal powders is a rapid prototyping technology in which a high-energy laser beam scans, melts, shrinks and consolidates metal powders with single component. For better understanding physical mechanisms during laser sintering of single-component metal particles, a temperature transforming model with the consideration of shrinkage and convective flows is introduced to analyze the thermal/fluid behaviors in selective laser sintering of single powder layer. The model is also applied to investigate the sintering of powders on top of existing sintered layers under single- multiple-line scanning manners according to the practical manufacturing processes.

## 1. Introduction

Selective laser sintering (SLS) of metal powders is a rapid prototyping technology that allows generating complex 3D parts by bonding powdered metal materials using a directed laser beam layer-by-layer (Kruth, 1991; Bourell et al., 1992; Agarwala et al., 1995). During the SLS process of metal, when the laser beam scans and melts a row of powder particles, the melted powder grains stick to each other via surface tension forces, thereby forming a series of spheres with diameters approximately equal to the diameter of the laser beam; this is referred to as "balling phenomena." To overcome the balling phenomenon, metal parts can be obtained by SLS of two-component powders, in which mixture of two different powders with significantly different melting points are laser sintered. To yield a fully-densified metallic part, the post-processing by liquid metal infiltration is often required (Agarwala, 1994).

As the ultimate goal of the rapid manufacturing processes is to fabricate functional parts directly from single material without the use of any intermediate binders and post-processing, there is increasing interest in the production of metallic objects via one-step process using single component powders. In the metal SLS of pure powders, scanning is carried out line by line and the energy of the laser causes melting along a row of powder particles, thereby forming a molten track of cylindrical shape. With the fast development of application of laser material processing especially in rapid prototyping /manufacturing technology, laser sintering of single-component metal powders for sealing relatively porous surfaces and/or achieving dense, homogenous surface layers has drawn increased interest over recent years (Tolochko et al., 2003).

The inherent complexity of this process under different conditions requires numerical methods to enable better understanding physical mechanisms during laser sintering of single-component metal particles. Melting and resolidification during metal SLS have significant effects on the temperature distribution during sintering. Convective flow in the molten pool strongly influences the energy transport, the propagation of the melting front during metal SLS process. A significant amount of numerical studies of laser processing of metal have been carried out by researchers in the past years (Kim et al., 1997; Jadi and Dutta, 2001; Chakraborty et al., 2004a, b; Li et al., 2004; Iwamoto et al., 1998; Han and Liou, 2004).

The thermal modeling of selective laser sintering of single-component powders (Xiao and Zhang, 2007a, b, 2008) will be presented in this paper. A temperature transforming model with consideration of shrinkage and convective flow during the process will be described. The model is also applied to investigate the selective laser sintering of metal powders on top of previously sintered layers under single- and multiple- line scanning manners.

## 2. Physical Models and Problem Statements

Figure 1(a) shows a schematic diagram for melting and resolidification of a loose metal powders layer under a moving Gaussian laser beam at a constant scanning velocity $u_b$ along the positive *x*-direction. Figure 1(b) shows the direct metal laser sintering process of a loose powder layer on top of multiple previously sintered layers under single-line laser scanning. Bonding occurs between laminates as the new layer is sintered. Figure 1(c) shows a schematic diagram of laser sintering of metal powders on top of sintered layers under multiple-line scanning. The laser beam scans the powder surface in a back-and-forth manner layer by layer. The overlaps not only between vertically deposited layers but also adjacent lines are considered. The overlap between adjacent lines is defined as the ratio of the width of the liquid pool at the bottom surface of the fresh layer and that of the liquid pool at the top surface.

The temperature transforming model using a fixed grid method is employed to describe the sintering problem. This model assumes that the solid-liquid phase change occurs over a range of melting temperature. Since the molten pool moves with the laser beam, the problem can be conveniently studied in a reference frame (X, Y, Z) fixed with the laser beam. The dimensionless governing equations in the moving coordinate system can be written as:

$$\frac{\partial U}{\partial X} + \frac{\partial V}{\partial Y} + \frac{\partial W}{\partial Z} = 0 \quad (1)$$

$$\frac{\partial U}{\partial \tau} + \frac{\partial [U(U - U_b)]}{\partial X} + \frac{\partial (UV)}{\partial Y} + \frac{\partial [U(W + W_s)]}{\partial Z}$$
$$= -\frac{\partial P}{\partial X} + \frac{\partial}{\partial X}\left(\Pr\frac{\partial U}{\partial X}\right) + \frac{\partial}{\partial Y}\left(\Pr\frac{\partial U}{\partial Y}\right) + \frac{\partial}{\partial Z}\left(\Pr\frac{\partial U}{\partial Z}\right) \quad (2)$$

$$\frac{\partial V}{\partial \tau} + \frac{\partial [V(U - U_b)]}{\partial X} + \frac{\partial (VV)}{\partial Y} + \frac{\partial [V(W + W_s)]}{\partial Z}$$
$$= -\frac{\partial P}{\partial Y} + \frac{\partial}{\partial X}\left(\Pr\frac{\partial V}{\partial X}\right) + \frac{\partial}{\partial Y}\left(\Pr\frac{\partial V}{\partial Y}\right) + \frac{\partial}{\partial Z}\left(\Pr\frac{\partial V}{\partial Z}\right) \quad (3)$$

$$\frac{\partial W}{\partial \tau} + \frac{\partial [W(U - U_b)]}{\partial X} + \frac{\partial (WV)}{\partial Y} + \frac{\partial [W(W + W_s)]}{\partial Z}$$
$$= -\frac{\partial P}{\partial Z} + \frac{\partial}{\partial X}\left(\Pr\frac{\partial W}{\partial X}\right) + \frac{\partial}{\partial Y}\left(\Pr\frac{\partial W}{\partial Y}\right) + \frac{\partial}{\partial Z}\left(\Pr\frac{\partial W}{\partial Z}\right) + \frac{\text{Ra}}{N_i}\Pr\theta \quad (4)$$

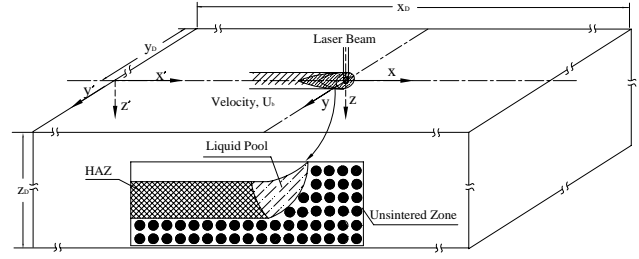

(a) Selective laser sintering of a single layer

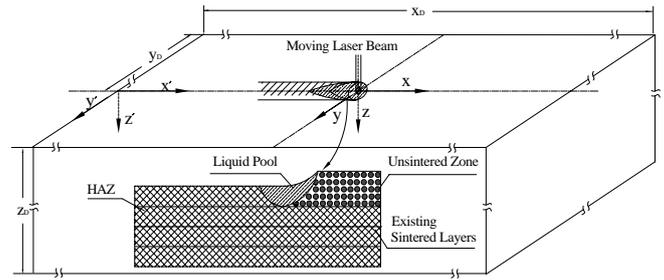

(b) Single-line scanning of loose powders on existing sintered layers

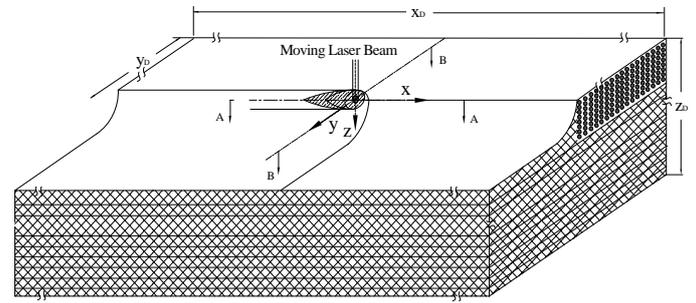

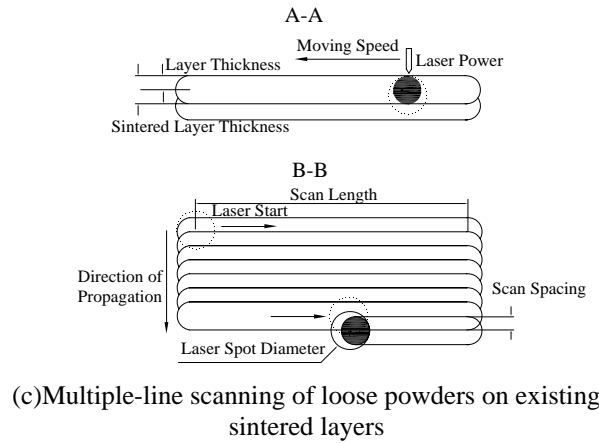

(c) Multiple-line scanning of loose powders on existing sintered layers

**Figure 1** Physical models

$$\frac{\partial(C\theta)}{\partial \tau}+\frac{\partial[C\theta(U-U_b)]}{\partial X}+\frac{\partial(C\theta V)}{\partial Y}+\frac{\partial[C\theta(W+W_s)]}{\partial Z}$$
$$=\frac{\partial}{\partial X}(K\frac{\partial \theta}{\partial X})+\frac{\partial}{\partial Y}(K\frac{\partial \theta}{\partial Y})+\frac{\partial}{\partial Z}(K\frac{\partial \theta}{\partial Z}) \quad (5)$$
$$-\left[\frac{\partial B}{\partial \tau}+\frac{\partial[(U-U_b)B]}{\partial X}+\frac{\partial(VB)}{\partial Y}+\frac{\partial[(W+W_s)B]}{\partial Z}\right]$$

where

$$W_s = \begin{cases} 0 & Z \geq S \\ \varepsilon\left(\frac{\partial S}{\partial \tau}-U_b\frac{\partial S}{\partial X}\right) & Z < S \end{cases} \quad (6)$$

$$\Pr = \begin{cases} N & \theta < -\delta\theta \\ \Pr_\ell + \frac{(\Pr_\ell - N)}{2\delta\theta}(\theta - \delta\theta) & -\delta\theta \leq \theta \leq \delta\theta \\ \Pr_\ell & \theta > \delta\theta \end{cases} \quad (7)$$

$$C = \begin{cases} C_{s\ell} & \theta < -\delta\theta \\ \frac{1}{2}(1+C_{s\ell}) + \frac{1}{2St\delta\theta} & -\delta\theta \leq \theta \leq \delta\theta \\ 1 & \theta > \delta\theta \end{cases} \quad (8)$$

$$K = \begin{cases} K_{eff} & \theta < -\delta\theta \\ K_{eff} + \frac{(1-K_{eff})}{2\delta\theta}(\theta + \delta\theta) & -\delta\theta \leq \theta \leq \delta\theta \\ 1 & \theta > \delta\theta \end{cases} \quad (9)$$

$$B = \begin{cases} 0 & \theta < -\delta\theta \\ \frac{1}{2St} & -\delta\theta \leq \theta \leq \delta\theta \\ \frac{1}{St} & \theta > \delta\theta \end{cases} \quad (10)$$

The boundary condition equation (5) written in dimensionless form is as follows,

$$-K\frac{\partial \theta}{\partial Z} = N_i \exp[-X^2 - Y^2]$$
$$-N_R\left[(\theta+N_t)^4-(\theta_\infty+N_t)^4\right]-B_i(\theta-\theta_\infty), \quad Z = S_0(X,Y) \quad (11)$$

The dimensionless form of Marangoni convection at the heating surface is

$$\frac{\partial V_s}{\partial n}+\frac{\partial V_n}{\partial s}=-\frac{Ma}{N_i}\frac{\partial \theta}{\partial s}, \quad Z = S_0(X,Y) \quad (12)$$

## 3. Numerical Solution

The sintering problem specified by equations (1)–(5) is a steady-state, three-dimensional, nonlinear problem. Since the locations of the solid-liquid interface and the heating surface are unknown *a priori*, a false transient method is employed to locate various interfaces. Steady-state solution is obtained when the temperature distribution and locations of various interfaces do not vary with the false time.

The governing equations were discretized and solved using the SIMPLE algorithm (Patankar, 1980). The convection and diffusion terms were discretized using the power-law scheme. A non-uniform 122×52×52 (in the X, Y, Z directions, respectively) grid number is used in the computation and the false time step is 0.001. The iterative procedure was continued until the following convergence criterion was satisfied:

$$\frac{\sum_P |\phi - \phi^{old}|}{\sum_P |\phi|} < 0.001 \quad (13)$$

where $\sum_P$ denotes summation overall grid points and $\phi$ is the variables being computed, e.g., $U$, $V$, $W$ and $T$.

## 4. Results and Discussions

Numerical calculations are performed in the three models for sintering of AISI 304 stainless steel powders.

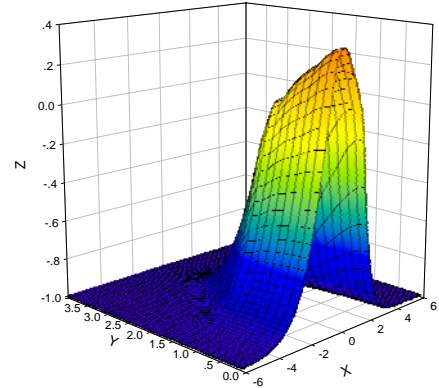

(a) Surface temperature distribution

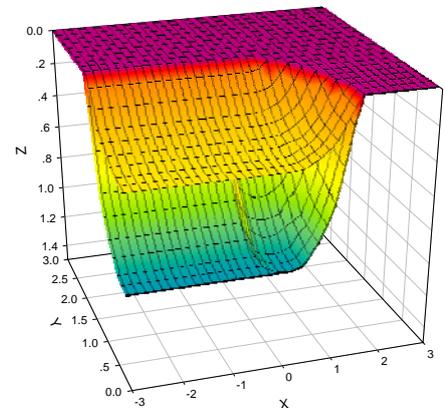

(b) Shape of interfaces

**Figure 2** Surface temperature and shape of interfaces ($N_i = 1.0$, $U_b = 0.05$ $\varepsilon = 0.4$, $Ra = 46.42$, $Ma = 2825.9$)

## 4.1 SLS of Single Powder Layer

Figure 2(a) illustrates the surface temperature distribution of the powder bed subjected to a moving laser beam. The peak temperature at the powder bed surface is near the trailing edge of the laser beam rather than at the center due to motion of the laser beam. Because the thermal conductivity in the molten pool is much larger than that in the unsintered zone, the temperature changes smoothly in the molten pool but sharply in the unsintered zone near the molten pool. Figure 2(b) shows the three-dimensional shape of the powder bed surface, molten pool and HAZ at the same condition of Fig. 2(a).

## 4.2 Single-Line scanning of loose powders on top of existing sintered layers

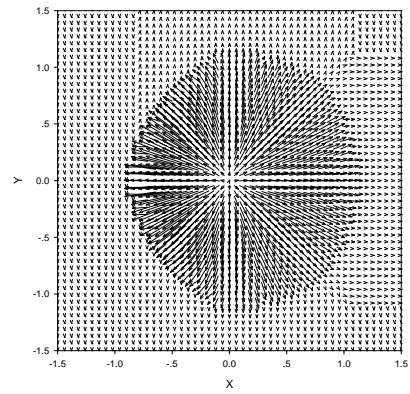

(a) Top view

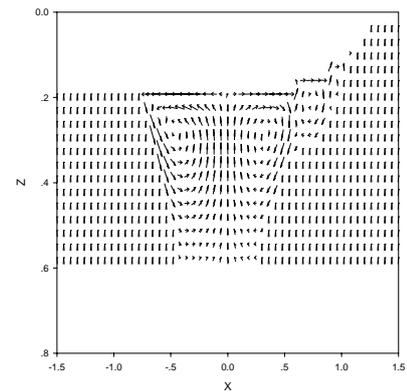

(b) Longitudinal view at y = 0

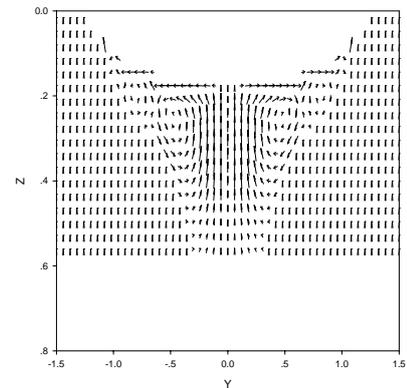

(c) Cross-sectional view at x = 0

**Figure 4** Dimensionless velocity vector plot
($\Delta_s = 0.4, U_b = 0.02, \varepsilon = 0.5, N = 5$)

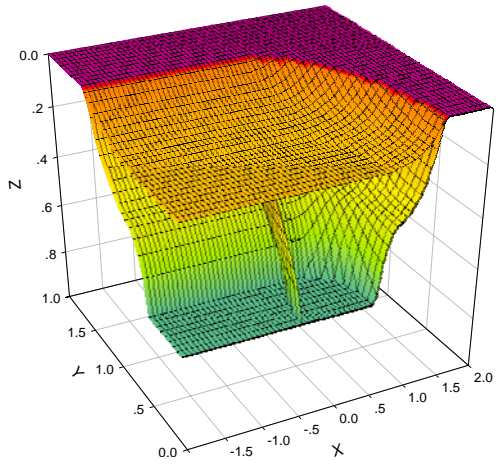

(a) $N = 1$

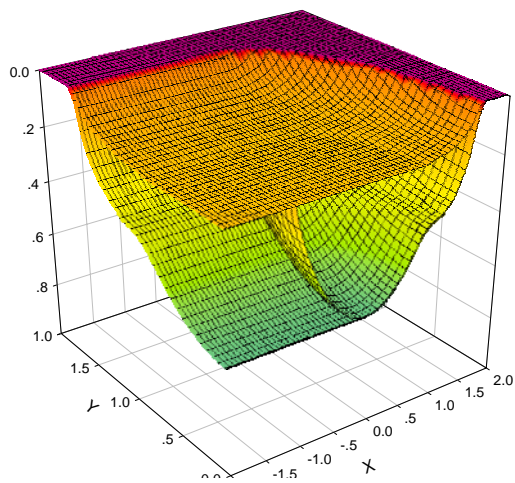

(b) $N = 5$

**Figure 3** Three-dimensional shape of the HAZ
($\Delta_s = 0.4, U_b = 0.02, \varepsilon = 0.5$)

Figure 3 shows the three-dimensional shapes of the powder bed surface, molten pool and HAZ when the number of sintered layers equals to 1 and 5 respectively under Single-Line scanning of loose powders on top of existing sintered layers. The depths of surface and HAZ decrease with the increasing X in the laser scanning direction Y.

Figure 4 show the velocity vectors in the liquid pool plotted in three different views ($\varepsilon = 0.5, N = 5$). Since the surface tension is a decreasing function of temperature, i.e., $\partial \gamma / \partial T$ is negative, the higher surface tension of the cooler liquid metal near the edge of the liquid pool tends to pull the liquid metal away from the center of the liquid pool, where the liquid metal is hotter and the surface tension is lower. Therefore, fluid flow on the surface of liquid pool is radially outward as can be seen in Fig. 4(a). The liquid metal flow is also driven by buoyancy force as illustrated in Fig. 4(b) and 4(c). The hotter liquid metal near the central region of the molten pool floats up to the surface, while the cooler liquid metal near the pool boundary sinks along the melt/solid interface to the bottom of the pool. This circulation of fluid flow induced by the surface tension gradient and buoyancy force is consistent with the typical natural convection pattern found in the literature.

### 4.3 Multiple-Line scanning of loose powders on top of existing sintered layers

Figure 5 show the three-dimensional shapes of the powder bed surface, molten pool and HAZ when $N=1$ and $N=3$ respectively ($N$ is the number of underneath existing sintered layer) under multiple-Line scanning of loose powders on top of existing sintered layers. The overlapped region has reached the bottom of the physical domain when the desired overlap between adjacent lines is achieved for $N=1$, thus makes the curve of the liquid pool is flat at the bottom. The shrinkage is most significant at the track of the laser beam center, and a groove shrinking surface is formed at the area around the laser beam. Because the thermal conductivity of the sintered region is much higher than that of the sintering region, the melting boundary in the Z direction at the sintering region changes faster than that at the sintered region.

Figure 6 illustrates the surface temperature distribution of the powder bed in the sintering process. The peak temperature at the powder bed surface is near the trailing edge of the laser beam rather than at the centre of the laser beam due to motion of the laser beam. Because the thermal conductivity in the molten pool is much larger than that in the unsintered zone, the temperature changes smoothly in the molten pool but sharply in the unsintered zone near the molten pool. And the surface temperature at the sintered part (y < 0) of the overlapped region is higher than that of the sintering powders (y > 0).

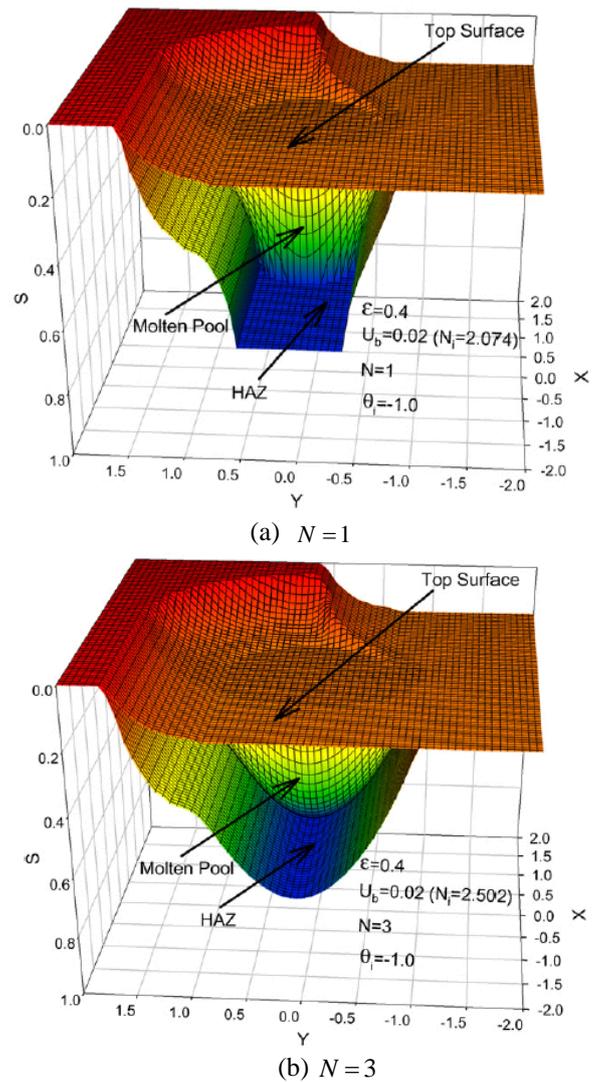

(a) $N=1$

(b) $N=3$

**Figure 5** Three-dimensional Shape of the Heat Affected Zone

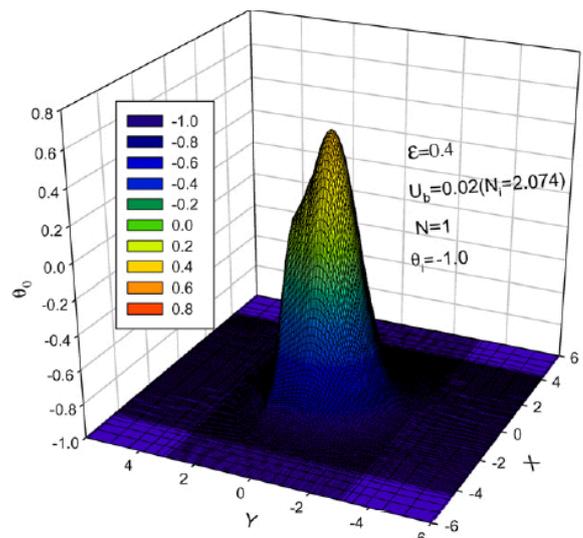

**Figure 6** Temperature distribution on the surface of the powder layer in multiple line scanning manner

## 5. Conclusion

A three dimensional numerical model for the convection-diffusion phase change process during laser sintering of single-component metal powders is presented. Temperature and velocity profile, as well as shapes of heating surface, liquid pool and heat affected zone were analyzed using temperature transforming model with the consideration of shrinkage and convective flow during the sintering processes. Direct metal laser sintering in the manufacturing processes under single-line and multiple-line scanning manners on top of sintered layers were also discussed.

## Acknowledgement

The work presented in this paper is supported by the U.S. Office of Naval Research under grant number N00014-04-1-0303.